\documentclass[manuscript, nonacm]{acmart}

\newcommand{\eventtype}{Workshop}
\newcommand{\event}{AI for Haptics and Haptics for AI: Challenges and Opportunities}

\acmBooktitle{2026 CHI Conference on Human Factors in Computing Systems (CHI '26), April 13--17, 2026, Barcelona, Spain}

\usepackage{blindtext}

\AtBeginDocument{%
  \providecommand\BibTeX{{%
    \normalfont B\kern-0.5em{\scshape i\kern-0.25em b}\kern-0.8em\TeX}}}

\begin{document}

\title{Why Modeling Human Haptic Material Perception with AI Is Difficult}
\author{Yasemin Vardar}
\affiliation{%
  \institution{Delft University of Technology (TU Delft)}
  \city{Delft}
  \state{South Holland}
  \country{The Netherlands}
}
\email{y.vardar@tudelft.nl}
\orcid{0000-0003-2156-1504}

\renewcommand{\shortauthors}{Vardar}

\begin{abstract}
Touch plays a central role in how humans perceive and recognize materials through physical contact. Despite decades of research, the mechanisms by which tactile signals are transformed into meaningful perceptual representations remain poorly understood, limiting the design of interactive systems and intelligent agents with human-like haptic perception. Recent advances in artificial intelligence (AI) offer new opportunities to model and exploit tactile data; however, haptics presents fundamental challenges for contemporary AI due to its interaction-dependent, multimodal nature. This position paper argues that progress at the intersection of AI and haptics is constrained by three key bottlenecks: (1) the scarcity of large, diverse, and balanced haptic datasets; (2) the lack of standardized evaluation platforms and perceptual benchmarks; and (3) limitations in model capacity and interpretability when applied to tactile perception. I discuss how these challenges impede generalization, reproducibility, and scientific insight into human touch and review emerging strategies to address them. This paper highlights opportunities for coordinated, cross-disciplinary efforts to advance AI systems that not only perform robust haptic perception but also contribute to a deeper understanding of human touch.
\end{abstract}

\keywords{Artificial Intelligence, Human Tactile Perception, Tactile Surfaces}
\begin{teaserfigure}
  \includegraphics[width=\textwidth]{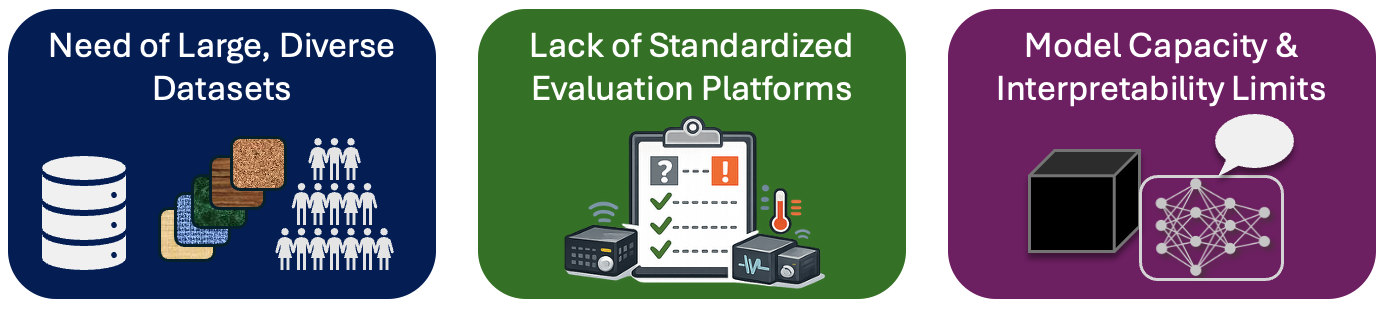}
  \caption{Illustration of challenges in modeling human haptic perception of materials outlined in this paper.}
  \Description{Three boxes each describing a challenge for human perception modeling: need for large, diverse datasets, lack of standardized evaluation platforms, and model capacity and interoperability limits.}
  \label{fig:teaser}
\end{teaserfigure}

\maketitle

\vspace{-0.7em}
\textbf{Reference Format:}\\
\begin{small}
\makeatletter \authors. \the\year. \@title. Proceedings of the \textit{\event} \eventtype\ at the \@acmBooktitle. \ref{TotPages}~pages. \makeatother
\end{small}

\section{Introduction}
When humans interact with materials through touch, a rich and diverse array of tactile signals is generated at the skin~\cite{johansson2009_review}. These signals convey information about materials’ sensory attributes—such as roughness, smoothness, hardness, and softness~\cite{okamoto2013_dimensions}—which shape perception and enable recognition. Despite decades of research, the mechanisms by which tactile information is rapidly transformed into meaningful perceptual attributes, and subsequently integrated into material recognition, remain poorly understood~\cite{richardson2022_learn2feel}. This limited understanding continues to constrain the design of interactive systems that meaningfully incorporate touch, as well as the development of robots and intelligent agents with human-like tactile perception.

Recent advances in artificial intelligence offer new opportunities to model haptic perception and to embed tactile intelligence into interactive systems. At the same time, touch poses fundamental challenges for contemporary AI approaches due to its inherently multimodal and interaction-dependent nature. Variability in exploratory behaviors, individual skin properties, and subjective perceptual judgments, combined with the need for specialized sensors and actuators, complicates data collection, representation, and evaluation. As a result, many AI methods that perform well in vision or language do not readily transfer to haptic perception.

This position paper argues that progress at the intersection of AI and haptics requires rethinking how tactile data are acquired, represented, and learned from, as well as how perceptual meaning emerges through physical interaction. In the context of the \textit{AI for Haptics, Haptics for AI} workshop, I outline key challenges and opportunities for building AI systems that both benefit from and contribute to a deeper understanding of human haptic perception.

\section{Need for Large, Diverse, and Balanced Datasets}
A central challenge in applying AI to haptic perception is the lack of large, diverse, and balanced datasets. In contrast to domains such as language, audio, and vision—where progress has been driven by massive datasets enabled by on-demand capture via standardized sensors and large-scale data sharing—tactile data acquisition requires active physical exploration with specialized sensing hardware. This process often demands expert knowledge, carefully designed protocols, and controlled laboratory conditions, making large-scale data collection costly and difficult.

Although the number of publicly available haptic datasets has grown in recent years~\cite{culbertson2014one, strese2017content, devillard2025_tactile, sens32025, toscani2022_viper}, most remain limited in scope. Many cover only a narrow range of materials, interaction types, or participants, and often focus on a single sensing modality such as vibration or friction. Other critical perceptual cues, such as thermal and softness, are frequently underrepresented. Even more comprehensive datasets, such as SENS3~\cite{sens32025}, are constrained by practical limits on material diversity, exploratory behaviors, and inter-participant variability. Because haptic perception emerges through physical contact, tactile signals are strongly shaped by exploration and the mechanical properties of the finger or tool~\cite{richardson2022_learn2feel, WIERTLEWSKI20121869,BALASUBRAMANIAN2026111054}. 

Capturing this variability is essential for generalization beyond laboratory settings.  Without it, AI models risk overfitting to narrow interaction regimes and failing to reflect the richness of human touch. Limited datasets can also produce models that reinforce incomplete or incorrect assumptions about human touch. Over-reliance on AI systems trained on narrow or sparse datasets can lead to unsafe or ineffective behavior in real-world applications. Moreover, the systematic underrepresentation of certain user populations, bodily characteristics, or sensory channels can exacerbate inequities and reduce accessibility.

Several strategies have been explored to address these limitations. One approach augments haptic data with more easily acquired modalities—such as images, video, or audio—to provide additional contextual information~\cite{jingu2025scene2hapcombiningllmsphysical, faruqi2025_tactstyle, cai2022_gan, ujitoko2018_image, sung2025_text2haptics, stroinski2025_text2haptics}. A second strategy focuses on modeling how variations in finger properties, tool characteristics, and exploratory behaviors~\cite{heravi2024_action} influence tactile signals, enabling adaptation and personalization in haptic interfaces. However, both methods remain immature and themselves require large, representative datasets for training and validation. A third emerging approach generates synthetic tactile data by linking signal properties to psychophysical features~\cite{hassan2020_authoring, begelinger2026_diffusion} or create new signals based on user preferences~\cite{lu2022_preference,zhang2025_texsense}, but this strategy is also vulnerable to modeling errors and perceptual biases. While promising, all of these approaches remain fundamentally constrained by the scarcity and limited diversity of available haptic datasets, which continues to represent a major bottleneck for progress in modeling human haptic perception via AI.

\section{Lack of Standardized Evaluation Platforms}

A further major challenge in applying AI to model human haptic perception is the absence of standardized evaluation platforms and metrics. In domains such as vision, audio, and language, progress has been driven not only by large datasets but also by the availability of widely adopted hardware and software infrastructures that reliably reproduce stimuli and support validation, annotation, and benchmarking. These shared platforms enable meaningful comparison across models, datasets, and studies.

In contrast, haptics lacks standardized devices capable of reproducing the full range of tactile sensations relevant to human perception~\cite{rodgriguez2025_weight}. Most haptic interfaces typically render only a limited subset of mechanical or thermal cues, and both output signals and evaluation metrics vary widely across devices and platforms~\cite{scheineder2022_sustainable}. As a result, experimental outcomes are often difficult to reproduce, and meaningful comparisons between systems or learning-based models remain challenging. This fragmentation creates a critical bottleneck: even with high-quality datasets and sophisticated AI models, it is difficult to reliably validate, benchmark, and deploy haptic systems in a reproducible manner.

Establishing multimodal hardware and generic evaluation frameworks is therefore essential for accelerating progress at the intersection of AI and haptics. Standardized benchmarks—whether through shared hardware, modular testbeds, or agreed-upon perceptual metrics—would enable consistent evaluation across studies and support the development of AI-driven haptic systems that generalize beyond individual devices or laboratories.

\section{Model Capacity and Interpretability Limitations}

While recent advances in machine learning have enabled increasingly expressive models for processing complex sensory data, haptic perception poses unique challenges related to both model capacity and interpretability. High-capacity models, such as deep neural networks, typically require large, diverse datasets to generalize effectively—conditions that are rarely met in haptics due to the cost, variability, and physical constraints of tactile data collection as summarized above. As a result, such models are prone to overfitting to specific sensors, interaction strategies, or experimental setups, limiting their robustness across users, materials, and contexts.

In parallel, many state-of-the-art models operate as black boxes, offering limited insight into how tactile signals are transformed into perceptual attributes or material representations. This lack of interpretability is particularly problematic in haptics, where perception emerges through embodied interaction and is closely linked to physical variables, such as force, motion, compliance, and contact geometry. Without interpretable representations, it becomes difficult to relate model behavior to established psychophysical findings, to diagnose failure modes, or to design haptic interfaces that leverage model outputs in perceptually meaningful ways.

Additionally, current learning approaches often conflate perceptual invariances with dataset-specific correlations. Models may achieve high performance within narrowly defined datasets while relying on incidental cues tied to particular sensors, exploratory behaviors, or participant populations. Such shortcut learning undermines generalization and limits the transferability of learned representations across devices and tasks. Addressing these issues requires models that can disentangle interaction dynamics, material properties, and perceptual attributes rather than implicitly encoding them.

\section{Conclusion}
In this position paper, I argued that while recent advances in artificial intelligence offer significant opportunities for modeling and leveraging haptic material perception, fundamental challenges remain. Unlike vision or language, haptic perception is inherently embodied, interaction-dependent, and highly prone to individual variability, placing unique demands on data collection, model design, and evaluation.

I highlighted three interconnected bottlenecks: the scarcity of large, diverse, and balanced haptic datasets; the lack of standardized evaluation platforms and metrics; and limitations in model capacity and interpretability when applied to tactile perception. Together, these challenges constrain the generalization and reproducibility of AI-driven haptic perception models and their translation into real-world systems. 

Addressing these challenges will require coordinated efforts across the AI and haptics communities. Promising directions include developing shared datasets and benchmarks, modular and standardized evaluation platforms, and learning approaches that integrate data-driven methods with physical, psychophysical, and embodied constraints. 

In this context, at the \textit{AI for Haptics, Haptics for AI} workshop, I propose to illustrate these challenges and present our recent works~\cite{sens32025, richardson2022_learn2feel, begelinger2026_diffusion, zou2026_decoding} as a basis for discussing opportunities and research directions for modelling human tactile perception of materials using AI. Through such efforts, my research group aims to develop AI systems that not only leverage tactile data to perform tasks but also provide insights into how humans perceive and interact with the world through touch.


\begin{acks}
This study was partially funded by the Dutch Research Council (NWO) under the VENI scheme (project number 19153) and the European Research Council (ERC) under the Starting Grant scheme (project number 101220242). 
\end{acks}

\bibliographystyle{ACM-Reference-Format}
\bibliography{mybibliography}

\section*{The Author}
Yasemin Vardar is an Associate Professor in the Department of Cognitive Robotics at TU Delft, The Netherlands, where she leads the Haptic Interface Technology Lab. She received her Ph.D. in Mechanical Engineering from Koç University, Turkey, in 2018, and worked as a postdoctoral researcher at the Max Planck Institute for Intelligent Systems in Germany. Her research focuses on understanding human touch and digitizing tactile information through advanced haptic interfaces. She is a recipient of the ERC Starting Grant, the NWO VENI Grant, and the Eurohaptics Best PhD Thesis Award, and currently serves as Chair of the IEEE Technical Committee on Haptics.

\end{document}